\documentclass[aps,pre,twocolumn,superscriptaddress,showpacs,tight]{revtex4-1}
\usepackage{amssymb}
\usepackage{epsfig}
\usepackage{amsmath}
\usepackage{times}
\usepackage{color}
\usepackage{lipsum}
\usepackage{ulem}
\usepackage{sidecap}

\begin{document}
\title{Spiral wave chimeras in reaction-diffusion systems: \\ phenomenon, mechanism and transitions}
\author{Bing-Wei Li}
\affiliation{Department of Physics, Hangzhou Normal University, Hangzhou 311121, China}
\author{Yuan He}
\author{Ling-Dong Li}
\author{Lei Yang}
\affiliation{Department of Physics, Hangzhou Normal University, Hangzhou 311121, China}
\author{Xingang Wang}
\email[Corresponding author. Email address: ]{wangxg@snnu.edu.cn}
\affiliation{School of Physics and Information Technology, Shaanxi Normal University, Xi’an 710119, China}

\begin{abstract} 
Spiral wave chimeras (SWCs), which combine the features of spiral waves and chimera states, are a new type of dynamical patterns emerged in spatiotemporal systems due to the spontaneous symmetry breaking of the system dynamics. In generating SWC, the conventional wisdom is that the dynamical elements should be coupled in a nonlocal fashion. For this reason, it is commonly believed that SWC is excluded from the general reaction-diffusion (RD) systems possessing only local couplings. Here, by an experimentally feasible model of three-component FitzHugh-Nagumo-type RD system, we demonstrate that, even though the system elements are locally coupled, stable SWCs can still be observed in a wide region in the parameter space. The properties of SWCs are explored, and the underlying mechanisms are analyzed from the point view of coupled oscillators. Transitions from SWC to incoherent states are also investigated, and it is found that SWCs are typically destabilized in two scenarios, namely core breakup and core expansion. The former is characterized by a continuous breakup of the single asynchronous core into a number of small asynchronous cores, whereas the latter is featured by the continuous expansion of the single asynchronous core to the whole space. Remarkably, in the scenario of core expansion, the system may develop into an intriguing state in which regular spiral waves are embedded in a completely disordered background. This state, which is named shadowed spirals, manifests from a new perspective the coexistence of incoherent and coherent states in spatiotemporal systems, generalizing therefore the traditional concept of chimera states. Our studies provide an affirmative answer to the observation of SWCs in RD systems, and pave a way to the realization of SWCs in experiments. \\ 

\end{abstract}

\date{\today }
\maketitle

\section{Introduction}

An intriguing phenomenon observed in systems of coupled identical oscillators is the coexistence of coherent and incoherent regions in the space, knowing as the chimera states \cite{kuramoto02,abrams04}. This counterintuitive dynamical behavior is discovered first by Kuramoto and Battogtokh \cite{kuramoto02}, and is named later as ``chimera state" for its analogy to the monster in Greek mythology which owns lion's head, goat's body, and serpent's tail \cite{abrams04}. Since its discovery, chimera state has inspired extensive theoretical and experimental studies during the past two decades, with the systems investigated ranging from physical to chemical and to biological systems \cite{MJP:2014,OEO:2008,GCS:2008,DMA:2008,IO:2011,IO:2013,MK:2013,YZ:2012,NS:2016,Tian_fop17,Xiao_nd18,Dai_ND18,Gavrilov_prl18,Xu_prl18,wang_epl19,Shepelev_cnsns19,QLD:2020,hagerstromnat12,tinsleynat12,nkomoprl13,martenspns13,Gambuzza_pre14,Larger_nc15,AEM:2020,Socorro_cnsns21}. With these studies, the strict conditions for generating chimera states as adopted in the seminal works have been largely relaxed \cite{sethiaprl14,Kemeth_chaos16,yeldesbayprl14,Laing_pre15,Bera_epl17,Tian_pre17}, and the concept of chimera state has been largely broadened and generalized \cite{YZ:2013,AZ:2014,Clerc_pre16,Bera_pre16_2,Premalatha_chaos18,shimapre04,martensprl10}. For instance, instead of nonlocal couplings which has been regarded as a necessary condition for generating chimera states, recent studies show that chimera states can also be generated in systems with global \cite{OEO:2008,sethiaprl14,yeldesbayprl14} or local couplings \cite{Laing_pre15,Bera_pre16,Clerc_pre16,Bera_epl17,Premalatha_chaos18,Kundu_pre18,Kundu_pre19,Clerca_cnsns20}; and, besides regular networks, a variety of chimera-like states have been reported and studied in networks of complex structures \cite{Yao_sr13,Zhu_pre14,XJ:2016,Majhi_plr19,LZH:2020,zheng_sspma20,Makarov_cnsns19}. In particular, chimera-like states have been observed in complex network of coupled neurons \cite{Majhi_plr19,LZH:2020}, and are regarded as having important implications to the neuronal functions, saying, for example, the unihemispheric slow-wave sleep (USWS) of some aquatic mammals (e.g. dolphins and whales) and birds \cite{NCR:2000}, in which one half of the brain is in sleep while the other part of the brain remains awake. 

Whereas chimera state is originally observed in one-dimensional systems, recent studies show that sophisticated chimera-like patterns can be also generated in higher dimensional systems. One example is the spiral wave chimera (SWC) \cite{shimapre04,martensprl10}, which combines the features of spiral waves and chimera states, and is typically observed in two-dimensional systems of nonlocally coupled oscillators. Different from the classical spiral wave in which the core (spiral tip) is defined as the point of phase singularity (topological defect), in SWC the core is constituted by a group of desynchronized oscillators and occupies a small, circular region in the space. Interestingly, it is shown that despite of the incoherent inner core, spiral wave is propagating stably in the outer region. Discovered by Shima and Kuramoto in 2004 in nonlocally coupled periodic oscillators \cite{shimapre04}, SWC has received growing interest in the field of nonlinear science in recent years, particularly for researchers working on pattern formations in reaction-diffusion (RD) systems \cite{laingphd09,YL:2011,guprl13,Wang_jcp14,xiepre15,Maistrenko,Li_pre16,Nicolaou_prl17,Omelchenko_siam18,Kundu_epj18,Gao_CSF18,Rybalova_chaos19,Totz_sr20,Maistrenko_epj20}. For the model of nonlocally coupled phase oscillators, an analytical description of SWC has been given in Ref. \cite{martensprl10} and, by the perturbation theory, both the rotation speed of the spiral arms and the size of the asynchronous core can be predicted. Besides phase oscillators, SWCs have been also observed in nonlocally coupled chaotic oscillators \cite{guprl13}, which are characterized by the presence of synchronization defect lines along which the local dynamics is periodic. Experimental verification of SWCs has been given in Ref. \cite{Totz_np18}, in which a large-size two-dimensional array of nonlocally coupled Belousov–Zhabotinsky (BZ) chemical oscillators are employed and some new dynamical features of SWCs are revealed, including the erratic motion of the asynchronous spiral core, the growth and splitting of the cores, and the transition from SWCs to incoherent states. Despite the progresses made, the mechanisms and properties of SWCs remain elusive and many questions remain not clear, e.g., the roles of the phase-lag parameter in generating SWCs \cite{martensprl10}, the transitions from SWCs to other states in the parameter space \cite{Totz_np18}, and the observability of SWCs in locally coupled systems \cite{Wang_jcp14,Li_pre16}.

For experimental physicists and chemists, a question of particular interest is whether SWCs can be generated in the general RD systems in which the dynamical elements are locally coupled through diffusions. Whereas results based on numerical simulations indicate that SWCs could be generated in locally coupled systems \cite{Wang_jcp14,Kundu_epj18}, the models employed in these studies seems somewhat artificial and are difficult to be realized in experiments. The experiment conducted by Totz {\it et al.} \cite{Totz_np18}, whereas is able to generate SWCs successfully, relies on the non-physical, nonlocal couplings that are realized through computer interface. As such, from the point of view of experimental studies, an urgent question to be answered is whether SWCs can be observed in the general RD systems possessing local, diffusive couplings. Should the answer be positive, the following-up questions are: (1) What are the properties of the SWCs? (2) How the SWCs are destabilized and transited to other states as the system parameters vary? and (3) Can the theoretical models be realized in experiments? In the present work, we attempt to address these questions by investigating the dynamics of an experimentally feasible RD system of local couplings. We are able to demonstrate that, even though the system elements are coupled locally, stable SWCs can still be generated in a wide region in the parameter space. We conduct a detailed numerical analysis on the properties of SWCs, and also the transitions of SWCs to other states in the parameter space. It is found that, while the SWCs share the properties of the conventional SWCs as observed in nonlocally coupled systems, they do possess some unique features, including the presence of SWCs in partial variables, the destabilization scenarios of SWCs, and the phenomenon of shadowed spirals. In particular, in shadowed spirals, regular spirals are emerged on top of the desynchronization background, which manifests from a new viewpoint the coexistence of coherence and incoherence in spatiotemporal systems, generalizing thus the concept of chimera states. Furthermore, treating the system as an ensemble of oscillators coupled through a common medium, we conduct a phenomenological analysis on the formation of SWCs, which provides insights on the mechanism of SWCs.  

The rest of the paper is organized as follows. In Sec. II, we will present the model of a three-component FitzHugh-Nagumo-type RD system, and describe the numerical methods used in simulations. In Sec. III, we will demonstrate the typical SWC states observed in simulations and, by the conventional approaches, characterize the properties of SWCs. In Sec. IV, we will propose the phenomenological theory, based on which the underlying mechanism of SWCs will be explored. In Sec. V, we will study the transition behaviors of SWCs in the parameter space, in which the two destabilization scenarios, namely core breakup and core expansion, will be discussed and the new phenomenon of shadowed spirals will be presented. Candidate experiments for verifying the theoretical findings will be given in Sec. VI, together with discussions and conclusion.

\section{Model and numerical methods}

Our model of locally coupled RD system reads \cite{Alonso_jcp11,Nicola_pre02,Li_pre16},
\begin{eqnarray}
\frac{\partial u}{\partial t} &=& \phi(au -\alpha u^{3}-bv -cw)+D_{u}\nabla^{2} u,\label{rd1} 
\\
\frac{\partial v}{\partial t}&=& \phi\epsilon_{1}(u - v)+D_{v}\nabla^{2} v,\label{rd2}
\label{rd} \\
\frac{\partial w}{\partial t} &=& \phi\epsilon_{2}(u - w) + D_{w}\nabla^{2} w, \label{rd3} 
\end{eqnarray}
which describe the dynamics of the concentrations of three chemical reactants, $u$, $v$ and $w$. This three-component RD system consists of a FitzHugh-Nagumo (FHN) kernel (consisting of $u$ and $v$) coupled to the third component $w$, and has been used in literature to investigate pattern formations in BZ systems dispersed in a water-in-oil Aerosol OT (AOT) microemulsion (BZ-AOT system) \cite{Cherkashin_jcp08} or to model spot dynamics in gas discharges \cite{Schenk_prl97}. In specific, in the BZ-AOT system $u$ is associated with the activator species ${\rm HBrO_{2}}$, while $v$ and $w$ represent the inhibitors ${\rm Br^{-}}$ and ${\rm Br_{2}}$, respectively \cite{Cherkashin_jcp08}. The parameters characterizing the local dynamics are $\phi$, $\alpha$, $a$, $b$, $c$, $\epsilon_{1}$ and $\epsilon_{2}$ (see Ref. \cite{Alonso_jcp11} for details). In specific, $a$ governs the reaction rate of $u$, and plays as the bifurcation parameter of the local dynamics; $\epsilon_2$ governs the reaction rate of $w$, through which the time-scale of $w$ can be tuned. In the present work, we fix the other parameters in the model, while investigating the variation of the system dynamics with respect to parameters $a$ and $\epsilon_{2}$, as $a$ and $\epsilon_{2}$ can be adjusted conveniently in experiments. The reactants diffuse in the space, with the coefficients being $D_u$, $D_v$ and $D_w$ for components $u$, $v$ and $w$, respectively. We note that the parameter $\phi$, which represents the fraction of the dispersed phase in the BZ-AOT system, in principle can be absorbed into the other parameters by a rescaling operation, but here we keep it in the equations so as to keep the model identical to the one studied in Ref. \cite{Alonso_jcp11}. 

Whereas diffusions exist for all chemical reactants, we focus on the case of single-component diffusion, i.e., $D_{u}=D_{v}=0$ and $D_w>0$. This setting is a good approximation for the realistic situations where the diffusion coefficient of one component is much larger than the others (e.g. $D_w>0$ and $D_{u}=D_{v}\approx 0$), and captures the essence of many chemical and biological systems in experiments, e.g., the synthetic genetic regulation network used in {\it Escherichia coli} cells \cite{Danino_nat10}, yeast cell layers \cite{Shutz_BJ11} as well as a dense population of {\it Dictyostelium} cells \cite{Noorbakhsh_pre15}. We note that with this setting, the RD system can be alternatively regarded as a population of FHN oscillators coupled through a dynamics environment described by $w$, namely the scheme of quorum sensing coupling \cite{JA:1976,AC:2006}. In this picture, the component $u$ of the local oscillators is affected by the environment through the term $-\phi c w$ in Eq. (\ref{rd1}), while the environment component $w$ is affected by the oscillators through the term $\phi \epsilon_{2}(u-w)$ in Eq. (\ref{rd3}). This picture of indirectly coupled FHN oscillators, as will be shown later, facilitates our analysis of the underlying mechanism of SWCs.  

In simulations, we employ the explicit forward Euler method to solve Eqs. (\ref{rd1}-\ref{rd3}), with the space step being $dx=dy=0.2$ and the time step being $dt=dx^2/(5D_{w})=0.016$. The two-dimensional spatially continuous RD system is discretized into a grid of $N_{tot}=N \times N$ oscillators, with $N=1024$. In implementing the Laplace term in Eq. (\ref{rd3}), a five-point stencil has been used. Throughout our present work, we fix the parameters $(\phi,b,c,\epsilon_1,\alpha,D_w) = (0.62,3.0,3.5,1.0,4/3,0.5)$, while varying the parameters $a$ and $\epsilon_{2}$ to change the system dynamics. The no-flux boundary condition is adopted in simulating the evolution of component $w$. As in other systems \cite{shimapre04,martensprl10}, special initial conditions are required in generating SWCs. Here, to generate SWCs, we initialize the systems with a SWC core, with the details the following. First, by simulating the dynamics of an isolated FHN oscillator [$D_u=D_v=D_w=0$ in Eqs. (1-3)], we obtain the time series of a component for one period of the oscillation, saying the series $u_{1},u_{2},u_{3},\cdots,u_{M}$, with $M$ the series length. Then, we assign each point on the grid with a data from the series $u_{j,k}=u_{[M\phi_{j,k}/(2\pi)]}$, with $(j,k)$ the location of the grid point, $\phi_{j,k}$ the geometry phase associated to $(j,k)$, and $[\cdot]$ the floor integral function. The geometry phase is defined as $\phi_{j,k}=\tan^{-1}(y_{j,k}-y_{c})/(x_{j,k}-x_{c})$, where $x_{j,k}=j$ and $y_{j,k}=k$ are the coordinates of the grid point $(j,k)$ in the two-dimensional space, and $(x_{c},y_{c})=(N/2,N/2)$ denotes location of the system center. The initial condition of the center point is set as $0$. Finally, we repeat the process for the other two components ($v$ and $w$), which completes the state initialization. 

\section{Spiral wave chimeras and properties}

\begin{figure*}[tbp]
\center
\includegraphics[width=0.75\linewidth]{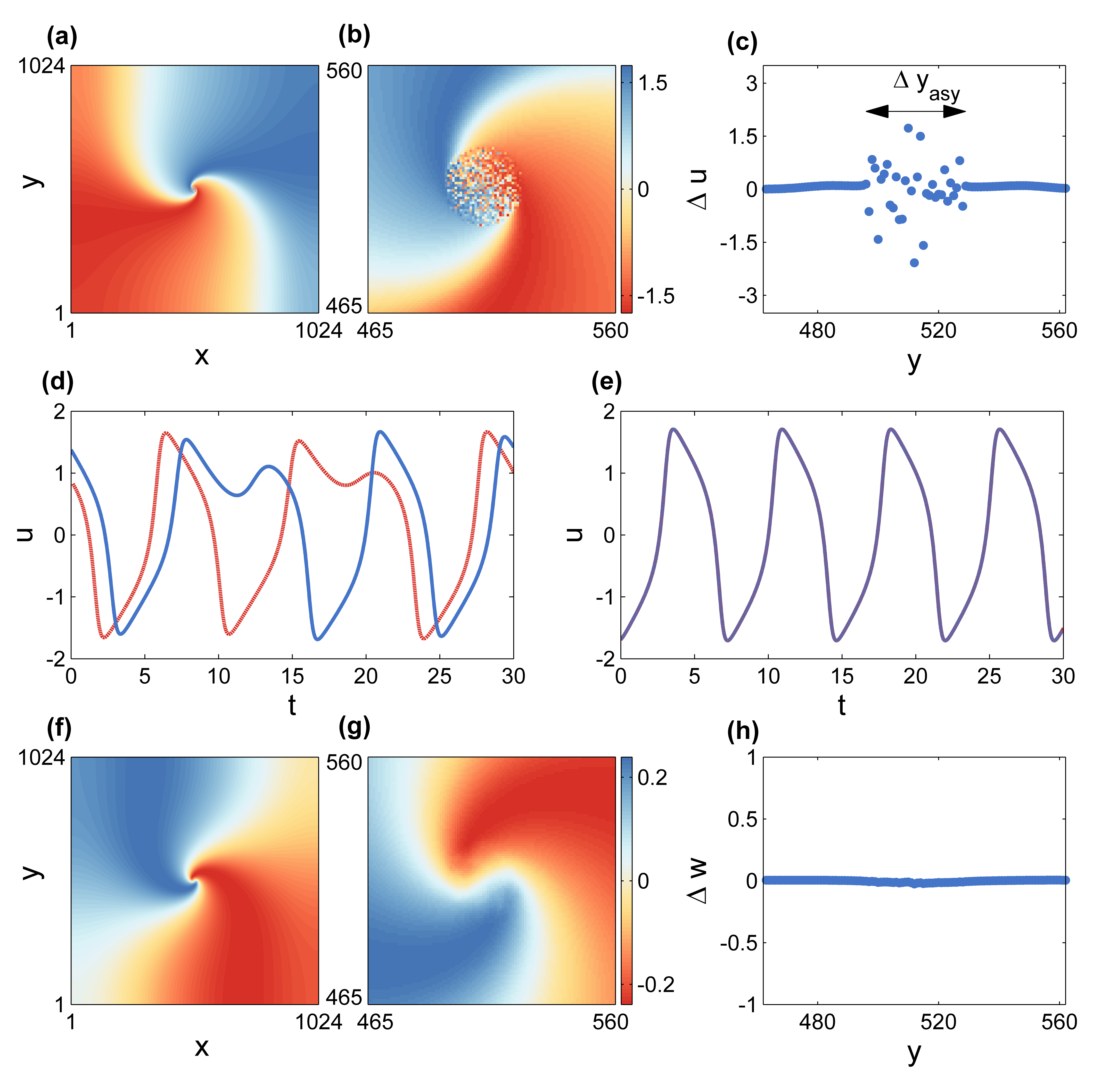}
\caption{For parameters $a=3.8$ and $\epsilon_{2}=0.2$, a typical SWC observed in simulations. (a) A snapshot of the $u$ component. (b) Enlarged view of the core region in (a). (c) The variation of $\Delta u(y) =u_{N/2,k+1}-u_{N/2,k}$ with respect to $y$ along the vertical central axis. The physical diameter of the asynchronous core is $d \approx \Delta y_{asy}\times dy \approx 33\times 0.2 =6.6$. (d) Time evolutions of two neighboring points inside the core region. (e) Time evolutions of two neighboring points outside the core region. (f) A snapshot of the $w$ component. (g) Enlarged view of the core region in (f). (h) The variation of $\Delta w(y)=w_{N/2,k+1}-w_{N/2,k}$ with respect to $y$ along the vertical central axis.}
\label{fig1}
\end{figure*}

Setting $a=3.8$ and $\epsilon_{2}=0.2$, we plot in Fig. \ref{fig1}(a) a snapshot of the $u$ component taken around $t=2\times 10^4$. We see that the whole space is occupied by a single spiral wave centered at $(0,0)$. As time increases, the spiral wave is rotating inwardly, e.g., towards the center, with the angular speed $\omega_s \approx 0.85$. A zoom-in plot of the core region is plotted in Fig. \ref{fig1}(b). We see that, encircled by the spiral arms, a circular region consisting of a group of disordered, asynchronous points is formed. (See Supplementary Materials for the movie.) To characterize the asynchronous core, we plot in Fig. \ref{fig1}(c) the variation of the reactant gradient $\Delta u=u_{N/2,k+1}-u_{N/2,k}$ with respect to $y$ along the vertical axis crossing the center. We see that $\Delta u$ is fluctuating randomly in the central region, $y\in (496,529)$, but is staying around $0$ outside. The physical diameter of the asynchronous core thus is estimated to be $d = \Delta y_{asy}\times dy=6.6$. To show the asynchronous feature of the core, we plot in Fig. \ref{fig1}(d) the time evolutions of two neighboring points inside the core, $(512,508)$ and $(512,507)$. Apparently, the two trajectories are desynchronized from each other. Figure \ref{fig1}(e) shows the time evolutions of two neighboring points adopted outside of core region, $(512,256)$ and $(512,255)$. We see that the two trajectories are completely overlapped. Similar to the component $u$, SWC is also observed for the component $v$ (not shown). Numerical results thus show that SWC can be generated in the locally coupled RD system.  

However, SWC is not observed for the third component $w$, as depicted in Figs. \ref{fig1}(f) and (g). In particular, Fig. \ref{fig1}(g) shows that in the central region the asynchronous core is disappeared. As such, the patten is evolving as a normal spiral. The absence of the asynchronous core is further verified by the gradient profile along the vertical central axis. As shown in Fig. \ref{fig1}(h), the value of $\Delta w=w_{N/2,k+1}-w_{N/2,k}$ is staying around $0$ in both the core and outer regions. The absence of SWC in the component $w$ is understandable, as $w$ is diffusing in the space with a fast speed ($D_w=0.5$), which smooths the distribution of $w$ in the core region. In contrast, as diffusion is absent for components $u$ and $v$, the asynchronous cores are stable. The coexistence of SWC (for components $u$ and $v$) and normal spiral (for the component $w$) is a unique feature for SWCs generated in locally coupled RD systems with single-component diffusion. 

\begin{figure*}[tbp]
\center
\includegraphics[width=0.75\linewidth]{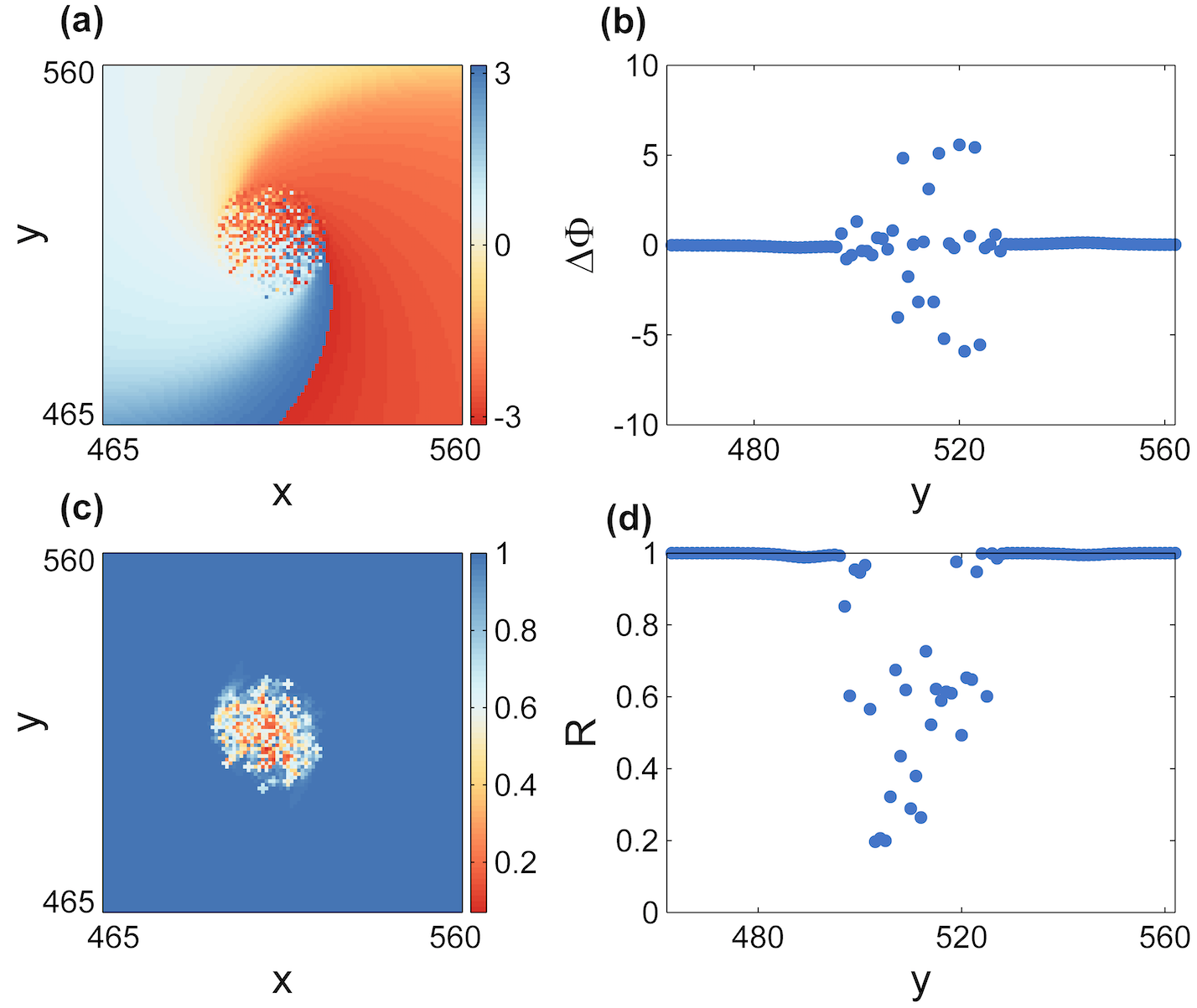}
\caption{Characterizing SWC by the phase variable and local order parameter. (a) The distribution of the phase variable $\Phi$ around the central region. (b) The variation of $\Delta \Phi(y)=\Phi_{N/2,k+1}-\Phi_{N/2,k}$ with respect to $y$ along the vertical central axis in (a). (c) The distribution of the local order parameter $R$ around the central region. (d) The variation of $R$ with respect to $y$ along the vertical central axis in (c). The SWC is the same to one shown in Fig. 1.}
\label{fig2}
\end{figure*} 

To study the properties of SWCs, we introduce the phase variable, $\Phi_{j,k} = \tan^{-1}(v_{j,k}/u_{j,k})$, and the local order parameter \cite{nkomoprl13}
\begin{equation}
R_{j,k}(t) = \left|\frac{1}{2m+1}\sum_{j',k'\in V_{j,k}}e^{{\rm i}\Phi_{j',k'}}\right|.
\end{equation}
Here ${\rm i}=\sqrt{-1}$ is the imaginary unit, and $V_{j,k}$ denotes the set of points around $(j,k)$ on the grid, including the point $(j,k)$ itself and its four nearest neighbors.
$m=2$ denotes the dimensionality of the RD system. It is straightforward to find that $R_{j,k}\approx 1$ if the set of points in $V_{j,k}$ are synchronized, and $0<R_{j,k}<1$ if the points are desynchronized. The spatial distribution of $\Phi_{j,k}$ around the core region is shown in Fig. \ref{fig2}(a), which is analogy to the SWC shown in Fig. \ref{fig1}(b). The phase difference $\Delta \Phi(y)=\Phi_{N/2,k+1}-\Phi_{N/2,k}$ along the vertical central axis is shown in Fig. \ref{fig2}(b). We see that, similar to the behavior of $\Delta u$ [see Fig. \ref{fig1}(c)], $\Delta \Phi$ is fluctuating randomly in the core region but is staying around $0$ outside. The distribution of the order parameter $R$ in the two-dimensional space and along the vertical central axis are presented in Figs. \ref{fig2}(c) and (d), respectively. We see that $R<1$ for points inside the core region, and $R\approx 1$ for the outside points, signifying the fact that points inside the core are desynchronized from their neighbors, while points outside the core are highly synchronized with their neighbors. 

We move on to characterize SWC by the topological charge \cite{Totz_sr20,Iyer_abe01} 
\begin{equation}
W = \frac{1}{2\pi} \oint_{C} \nabla \Phi \cdot d\vec{s},
\label{charge}
\end{equation}  
with $\Phi$ the phase variable defined above and $C$ a closed curve surrounding the asynchronous core. Previous studies show that for the case of single SWC (e.g. one asynchronous core and one spiral), the integral gives $\pm 2\pi$, resulting in $W=\pm 1$, with the sign of $W$ denoting the chirality of SWC.  For the SWC shown in Figs. \ref{fig1} and \ref{fig2}, we have $W=-1$, indicating that the SWC is left-handed \cite{Iyer_abe01}. When multiple SWCs exist, the net charge of the system will be keeping unchanged during the process of system evolution, which will be discussed later in exploring SWC transitions.

\begin{figure}[tbp]
\center
\includegraphics[width=0.9\linewidth]{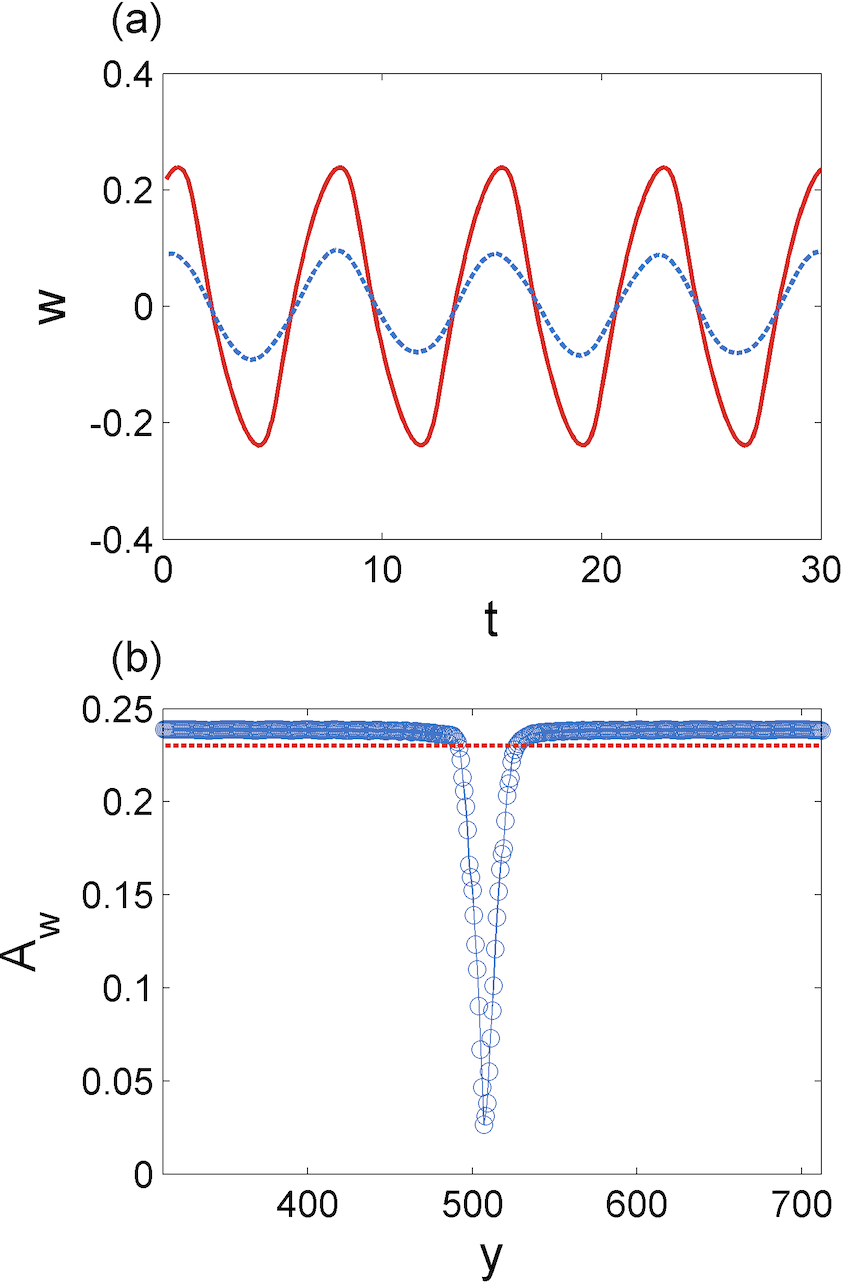}
\caption{Properties of the component $w$. (a) Time profile of $w$ for two grid points, one is far from the core region (red, solid line) and the other is inside the core (blue, dotted line). (b) Time-averaged amplitude, $ A_{w}(y)$, along central vertical axis ($x=512$). $A_w^c\approx 0.24$ denotes the critical amplitude for synchronization, below which the oscillator is not locked to the medium. The parameters are the same as in Fig. 1.}
\label{fig3}
\end{figure}

\section{Mechanism analysis}\label{mechanism}

The fact that SWCs can be generated in locally coupled RD systems seems contradictory to the existing studies on spiral waves, as it is well known that the presence of diffusion in RD systems will lead to a smooth distribution of the reactants in space (except the point at the spiral tip) \cite{Spiral:Book}. The key to generating SWCs in our model of locally coupled RD systems lies in the special scheme of single-component diffusion, i.e., diffusion exists only for the component $w$, while are absent for components $u$ and $v$. Such a scheme sets a barrier between the diffusive and non-diffusive components, protecting therefore the asynchronous cores from destruction. Indeed, as depicted in Fig. \ref{fig1}, the behaviors of the diffusive ($w$) and non-diffusive ($u$ and $v$) components are clearly different from each other. This might explain why SWCs have not been reported in locally coupled RD systems in literature.  

The mechanism of SWCs can be analyzed by a phenomenological theory, as follows. We first note that the RD system described by Eqs. ({\ref{rd1}-\ref{rd3}}) can be treated as a population of FHN oscillators (each has two variables, $u$ and $v$) coupled indirectly through a common medium (described by $w$) \cite{Li_pre16,cao_chaos19,li_jnls17}. In this picture, the local oscillators are isolated from each other, but are all driven by a spatially extended dynamical medium. Denoting $F(w_y,t)=-\phi c w_y(t)$ as the driving force at $y$ ($x=N/2$), the dynamics of the oscillator at $y$ is governed by the equation
\begin{eqnarray}\label{meanfield}
\frac{d u_y}{d t} &=& \phi(au_y -\alpha u_y^{3}-bv_y )+F(w_y,t), 
\\
\frac{d v_y}{d t}&=& \phi\epsilon_{1}(u_y - v_y) \label{mf-2}.
\end{eqnarray}
We note that the driving force $F(w_y,t)$ depends on both the spatial location of the oscillator and time, i.e., its a spatiotemporal signal. For the reason that the dynamics governing $w$ is linear [see Eq. (\ref{rd3})] and $u$ is oscillatory [see Fig. \ref{fig1}], the local component $w_y$ will be also oscillating with time. The oscillatory feature of $w_y$ is confirmed by simulations, as depicted in Fig. \ref{fig3}(a). It is noticed in Fig. \ref{fig3}(a) that the two oscillations, one in the core region and the other one in the outer region, are of similar frequency but different amplitudes. Specifically, the amplitude of the inner point is clearly smaller than that of the outer point. Denote $A_w(y)$ as the time-averaged amplitude of the oscillation at $y$, we plot in Fig. \ref{fig3}(b) the variation of $A_w(y)$ with respect to $y$ along the vertical central axis (i.e. $x=N/2$ in the pattern). We see that $A_w(y) \approx 0.24$ in the outer region, but is gradually decreased in the core region as $y$ approaches the central point. In particular, the value of $A_w(y)$ is decreased to about $0.02$ at $y=N/2$. 

\begin{figure}[tbp]
\center
\includegraphics[width=0.9\linewidth]{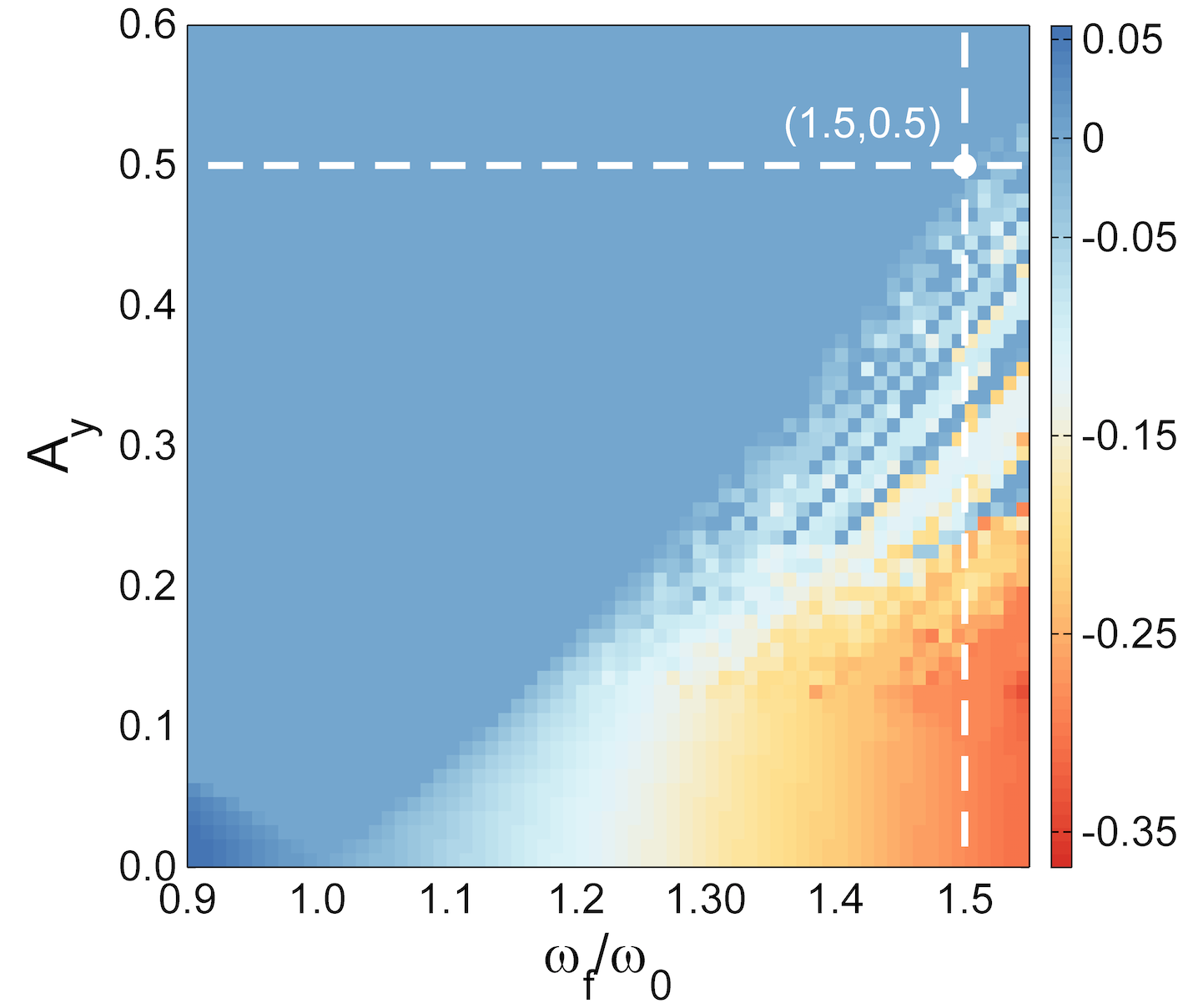}
\caption{Driving FHN oscillator by the periodic forcing, $F(w_y,t)=A_y\sin(\omega_{f}t)$, the distribution of the frequency error, $\Delta \omega$, in the two-dimensional parameter space spanned by the frequency ratio, $r=\omega_f/\omega_0$, and the driving amplitude, $A_y$. Vertical dashed line at $r=1.5$ corresponds to the frequency ratio of the SWC shown in Fig. \ref{fig1}. Along the vertical dashed line, we have $\Delta \omega \approx 0$ for $A_y>A_c\approx 0.5$.}
\label{figadd}
\end{figure}

Based on the numerical results [Fig. \ref{fig3}(a)], we may approximate the oscillations of $w_y(t)$ by a sinusoidal function, $w_y(t)=A_w(y) \sin(\omega_f t)$. Accordingly, the driving force can be written as $F(w_y,t)=A_y\sin(\omega_f t)$, with $A_y=-\phi c A_w(y)$. With this approximation, we are able to analyze the formation of SWC by a phenomenological approach, with the details the following. According to the synchronization theory \cite{pikovsky:book}, periodic oscillator of natural frequency $\omega_0$ can be locked to the external forcing given that the frequency mismatch between them is small and the driving amplitude is large enough. For the isolated FHN oscillator adopted in our study, the natural frequency is about $\omega_0=0.57$. Defining the frequency ratio $r = \omega_f/\omega_0$ and characterizing synchronization degree by the frequency error, $\Delta \omega=\Omega_0 - \omega_f$ with $\Omega_0$ the practical frequency of the FHN oscillator, we calculate numerically the distribution of $\Delta \omega$ in the two-dimensional parameter space spanned by $r$ and $A_y$. The results are plotted in Fig. \ref{figadd}. We see that with the increase of $r$, the critical amplitude, $A_{c}$, required for phase synchronization is gradually increased. As $\omega_f \approx 0.85$ (the rotating frequency of the spiral arms), the frequency ratio therefore is $r\approx 1.5$, which, according to the numerical results shown in Fig. \ref{figadd}, gives the critical amplitude $A_{c}\approx 0.5$. From the relation $A_w=A_{y}/(c\phi)$ and setting $A_{y}=A_{c}$, we finally have the critical amplitude $A_w^c\approx 0.23$, which defines the boundary of the asynchronous core. As depicted in Fig. \ref{fig3}(b), this predication is in good agreement with the results of direct simulations where $A_{w}\approx 0.24$ in the outer region. 

The above mechanism provides insights on the necessary conditions for generating SWC in locally coupled RD systems. Our above analysis shows that the nature of single-component diffusion is to weaken the synchronization of the local oscillators, so as to form the asynchronous core. Otherwise, if diffusions are introduced to all three components, the local oscillators will be strongly coupled and be synchronized, therefore destroying the asynchronous core. As such, the essence for generating SWC is that the local oscillators should be weakly coupled, instead of the scheme of single-component diffusion. This new understanding has been verified by simulations. For instance, if weak diffusions are introduced to component $u$ and $v$ ($D_u,D_v\ll D_w$), the SWC shown in Fig. 1 can still be observed (not shown). 

\section{Transitions from spiral wave chimeras to other states}

By varying the system parameters, the system may transit from SWC to other states. In the current study, we focus on the transitions of SWCs to other states with respect to the variations of $a$ and $\epsilon_{2}$. As discussed in Sec. II, $a$ plays as the bifurcation parameter of FHN oscillator and $\epsilon_2$ characterizes the reaction rate of the diffusive component $w$. We thus expect that by varying $a$ and $\epsilon_2$, rich dynamics could be observed. In what follows, we will present two typical scenarios governing the transitions: core breakup and core expansion. The former is featured by the continuous breakup of the asynchronous core, by which the system is finally developed into SWC turbulence. The latter, on the other hand, is featured by the continuous expansion of the asynchronous core, by which the system might develop into a new type of chimera state, the shadowed spirals.

\begin{figure*}[tbp]
\center
\includegraphics[width=0.8\linewidth]{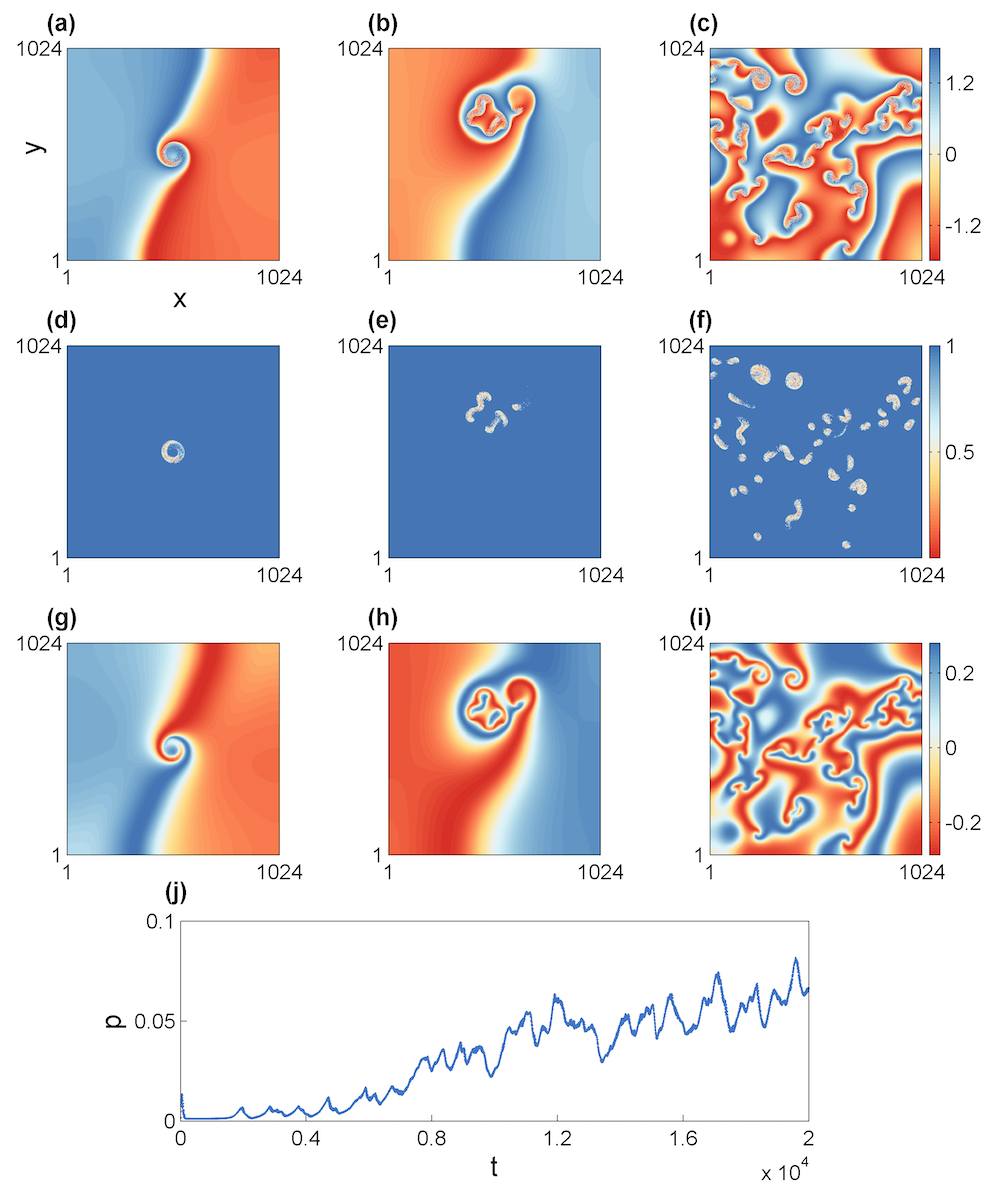}
\caption{ Transition from SWC to SWC turbulence through core breakup. Shown are the typical states of the component $u$ (a-c), the local order parameter $R$ (d-f) and the component $w$ (g-i) observed in the process of system evolution at $t=2000$, $6000$ and $20,000$. (j) Time evolution of the fraction of asynchronous oscillators, $p$, in the system. The parameters are $a=3.8$ and $\epsilon_{2}=0.245$.}
\label{fig5}
\end{figure*}

\begin{figure*}[tbp]
\center
\includegraphics[width=0.85\linewidth]{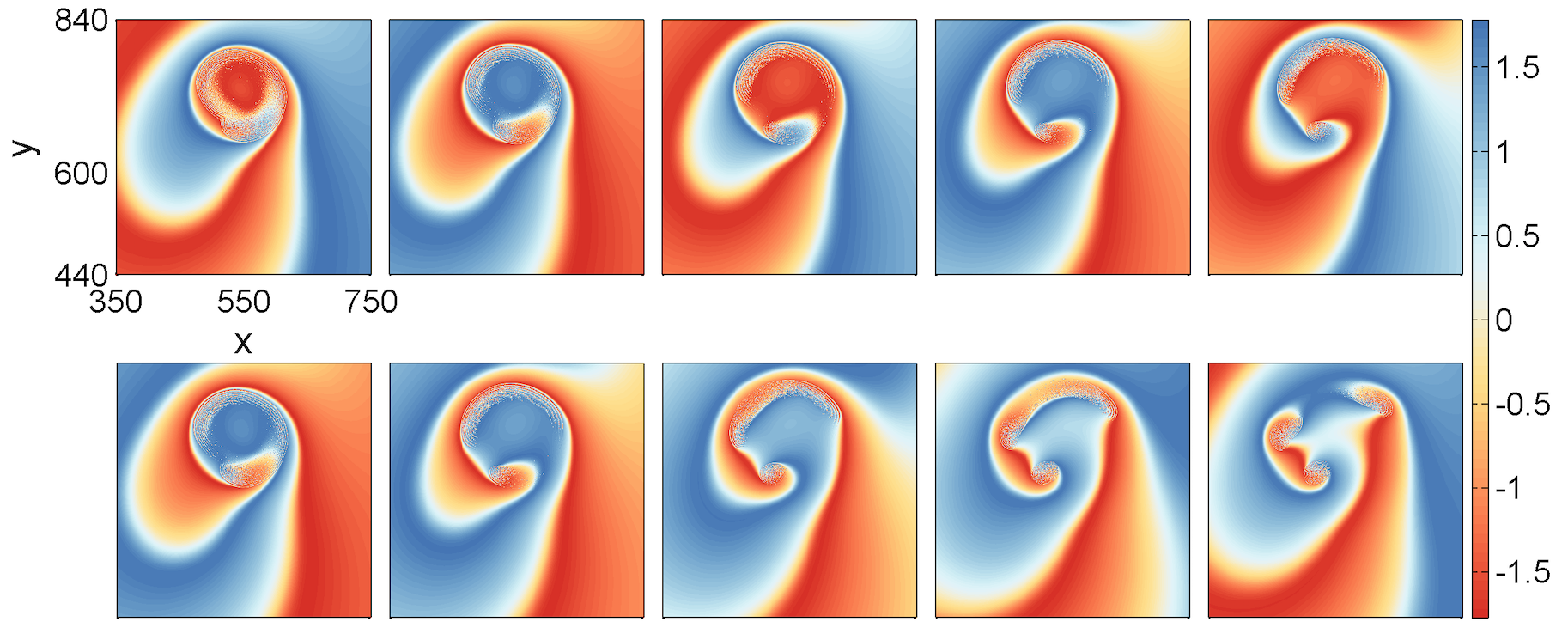}
\caption{ Details of core breakup. (a-j) Successive snapshots of the component $u$, with time interval being $\Delta t=32$. The parameters are the same as in Fig. \ref{fig5}.}
\label{fig6}
\end{figure*}

\subsection{Core breakup and SWC turbulence}

To demonstrate the scenario of core breakup, we fix the parameter $a=3.8$, while increasing $\epsilon_{2}$ gradually from $0.2$ (the same parameter and initial condition as used in Fig. \ref{fig1} for generating SWC). Numerical results show that when the increment is small, SWC survives, but with the wave length being slightly decreased. The SWC, however, becomes unstable when $\epsilon_{2}$ exceeds a critical value $\epsilon^c_{2} \approx 0.24$. To show an example, we set $\epsilon=0.245$ and plot in Fig. \ref{fig5} the typical states observed in the system evolution. The time evolution of the component $u$ is plotted in Fig. \ref{fig5}(a-c). We see that as time increases, the single asynchronous core is broken into many small-size asynchronous cores, leading to a state similar to spiral wave turbulence \cite{Spiral:Book}. However, different from the conventional picture of spiral wave turbulence, here the tips of the small spirals are replaced by asynchronous cores. For this reason, we call this new state SWC turbulence. The time evolution of the local parameter $R$ are plotted in Fig. \ref{fig5}(d-f), which show clearly how new asynchronous cores are born with the vanishing of the original core. The similar phenomenon is also observed for component $w$ [Fig. \ref{fig5}(g-i)]. As asynchronous core is absent in $w$, the state should be classified as spiral wave turbulence. To explore further the transition from SWC to SWC turbulence, we plot in Fig. \ref{fig5}(j) the time evolution of the fraction of asynchronous oscillators in the system, $p=N_{asy}/N_{tot}$ \cite{Kemeth_chaos16}. Here $N_{asy}$ is the number of oscillators whose local parameter is smaller to a threshold. For illustration, we set the threshold as $R=0.95$. Figure \ref{fig5}(j) shows that $p$ starts to increase at about $t=4800$, indicating the breaking of the asynchronous core at this moment. After that, $p$ is gradually increased with fluctuations, signifying the fact that more asynchronous cores are emerged and the SWCs are evolving as turbulence \cite{Kemeth_chaos16}. 

To have a close look at the transition dynamics, we focus on the behavior of the asynchronous core at the onset of the breaking ($t\approx 4800$). Typical states of $u$ observed during the breaking process are shown in Fig. \ref{fig6}. We see that the breaking starts with the emergence of a synchronous core inside the asynchronous core [see the region indicated by the white arrow in Fig. \ref{fig6}(a)]. As time increases, the synchronous core expands in space and, as the consequence, the asynchronous core is reshaped into an annulus [Fig. \ref{fig6}(b)]. The expansion of the synchronous core eventually leads to the breaking of the asynchronous annulus, resulting in two disconnected asynchronous segments [Fig. \ref{fig6}(c)]. In the following evolution, the asynchronous segments are pushed outward by the expanding synchronous core [Fig. \ref{fig6}(d)], and reshaped continuously by the rotation of the spiral arms [Figs. \ref{fig6}(e-i)]. Finally, with the insertion of the spiral arm, the synchronous core is broken into two parts [Fig. \ref{fig6}(j)]: the one connected with the spiral arms is developed to a small-size SWC, while the one detached from the arms is eventually developed into two small-size SWCs. (See Supplementary Materials for the movie.)

The above process of core breaking continues, resulting in SWC turbulence in which many small-size SWCs coexist, which, as the system evolves, are continuously broken and eliminated. It is just the breaking and elimination of the small SWCs that results in the wild fluctuation of $p$, as depicted in Fig. \ref{fig5}(j). To characterize SWC turbulence, we calculate the network topological charge of the system and investigate its time evolution. Previous studies on spiral wave turbulence show that, despite the continuous breaking and elimination of the spirals, the net topological charge is keeping unchanged \cite{Spiral:Book}. For the case of SWC breaking, this is indeed what we find in simulations. For instance, for the breaking process shown in Fig. \ref{fig6}, the breaking of the original core finally leads to the generation of three new cores [Fig. \ref{fig6}(j)]. The one connected to the spiral arms has the topological charge $W=-1$, while the other two detached from the arms have the charge $W=+1$ and $W=-1$. As such, the net topological charge of the system is keeping as $-1$. This finding is consistent with the results reported in Ref. \cite{Totz_np18}, in which splitting of SWCs in a nonlocally coupled oscillator system has been studied.  

\subsection{Core expansion and shadowed spirals}

\begin{figure*}[tbp]
\center
\includegraphics[width=0.75\linewidth]{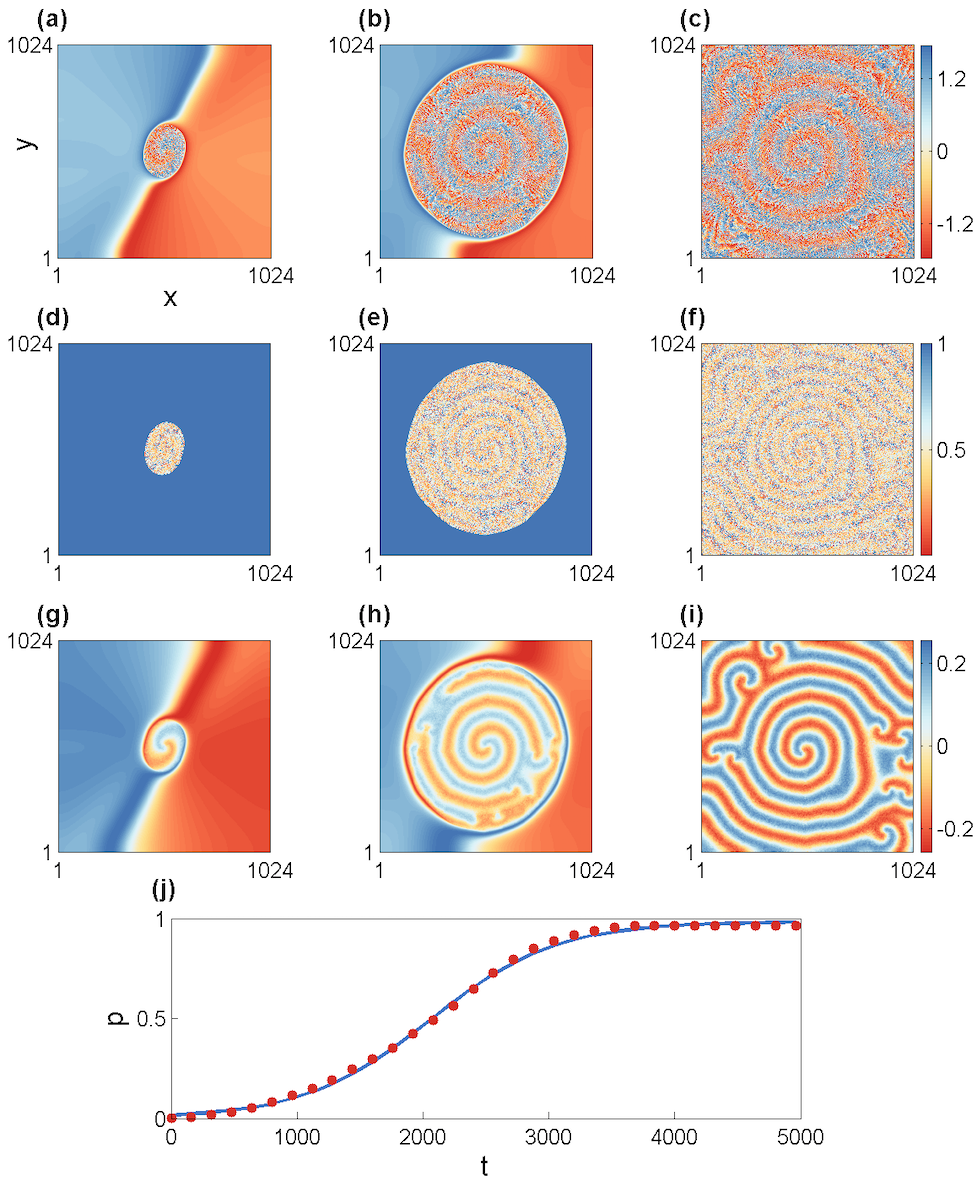}
\caption{Emergence of shadowed spirals through core expansion. Shown are the typical states of the component $u$ (a-c), the local order parameter $R$ (d-f) and the component $w$ (g-i) observed in the process of system evolution at $t=500$, $2000$ and $5000$. (j) Time evolution of the fraction of asynchronous oscillators, $p$, in the system. The numerical results (red dots) can be fitted by the logistic growth (blue curve). The parameters used in simulations are $a=3.8$ and $\epsilon_{2}=0.45$.}
\label{fig7}
\end{figure*}

Besides the scenario of core breakup, SWC may also be destabilized through the scenario of core expansion, which occurs when $\epsilon_{2}$ is large. Setting $\epsilon_{2}=0.45$, we plot in Figs. \ref{fig7}(a-c) three typical states observed in the evolution of $u$. We see that as time increases, the asynchronous core is gradually expanded and finally dominates the whole space. More interestingly, inside the expanding core, a distinct pattern of spiral waves is emerged on top of the disordered, desynchronization background. (See Supplementary Materials for the movie.) The feature of core expansion is also reflected in the evolution of the local parameter $R$, as can be seen in Figs. \ref{fig7}(d-f). Moreover, in Fig. \ref{fig7}(f) it is clearly seen that the local order parameter is close to $1$ for sites outside the core region, while is smaller to $1$ for sites inside the core. For the reason that the spiral waves are embedded in the background of desynchronized sites, we call this new phenomenon shadowed spirals, so as to distinguish it from the conventional spirals observed in RD systems. The development of new spiral waves is more clearly presented in the component $w$ [Figs. \ref{fig7}(g-i)], in which the desynchronization background is disappeared and only the spirals are shown. To explore the transient behavior of the core expansion, we plot in Fig. \ref{fig7}(j) the time evolution of the fraction of asynchronous oscillators, $p$. Different from the core breakup scenario [see Fig. \ref{fig5}(j)], here we see that $p$ is smoothly increased with time, and reaches $1$ at about $t=4\times 10^3$, indicating that at this moment the whole space is dominated by shadowed spirals. Numerically, we find that the behavior of $p$ can be fitted by the logistic growth
\begin{equation}
p(t)=\frac{\beta}{1+\gamma \exp(-kt)},
\end{equation}
with the fitted coefficients being $\beta=0.9836$, $\gamma=6.1$ and $k=2\times 10^{-3}$. 

\begin{figure*}[tbp]
\center
\includegraphics[width=0.8\linewidth]{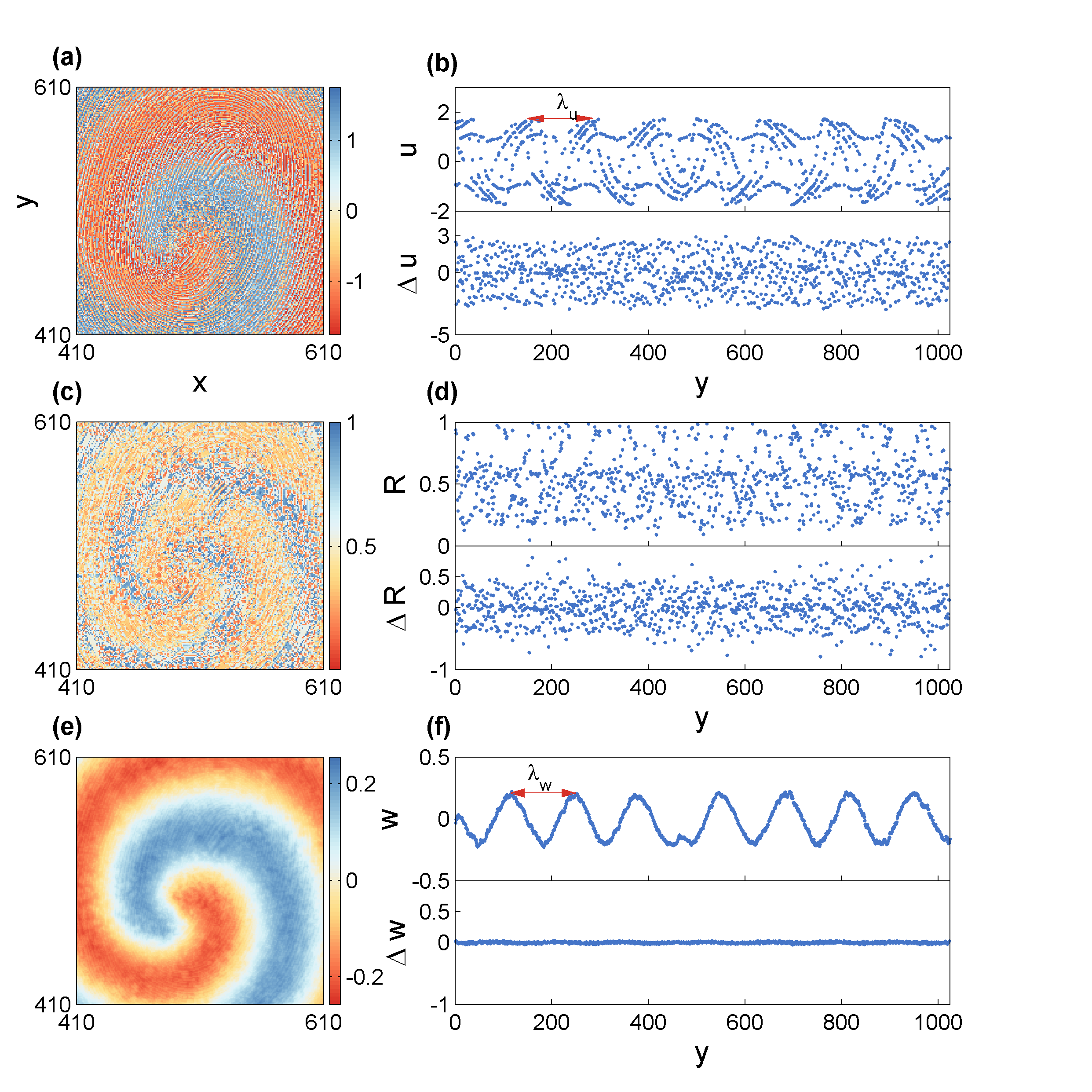}
\caption{Properties of shadowed spirals in the core area. (a) Enlarged view of component $u$. (b) Distributions of $u$ (top) and $\Delta u$ (bottom) along the central vertical axis ($x=512$) in (a). The wave length of $u$ is $\lambda_u\approx 26$. (c) Enlarged view of the local order parameter $R$. (d) Distributions of $R$ (top) and $\Delta R$ (bottom) along the central vertical axis in (c). (e) Enlarged view of component $w$. (f) Distributions of $w$ (top) and $\Delta w$ (bottom) along the central vertical axis in (e). The wave length of $w$ is $\lambda_w\approx 26$. The parameters are the same as in Fig. \ref{fig7}.}
\label{fig8}
\end{figure*}

\begin{figure*}[tbp]
\center
\includegraphics[width=0.8\linewidth]{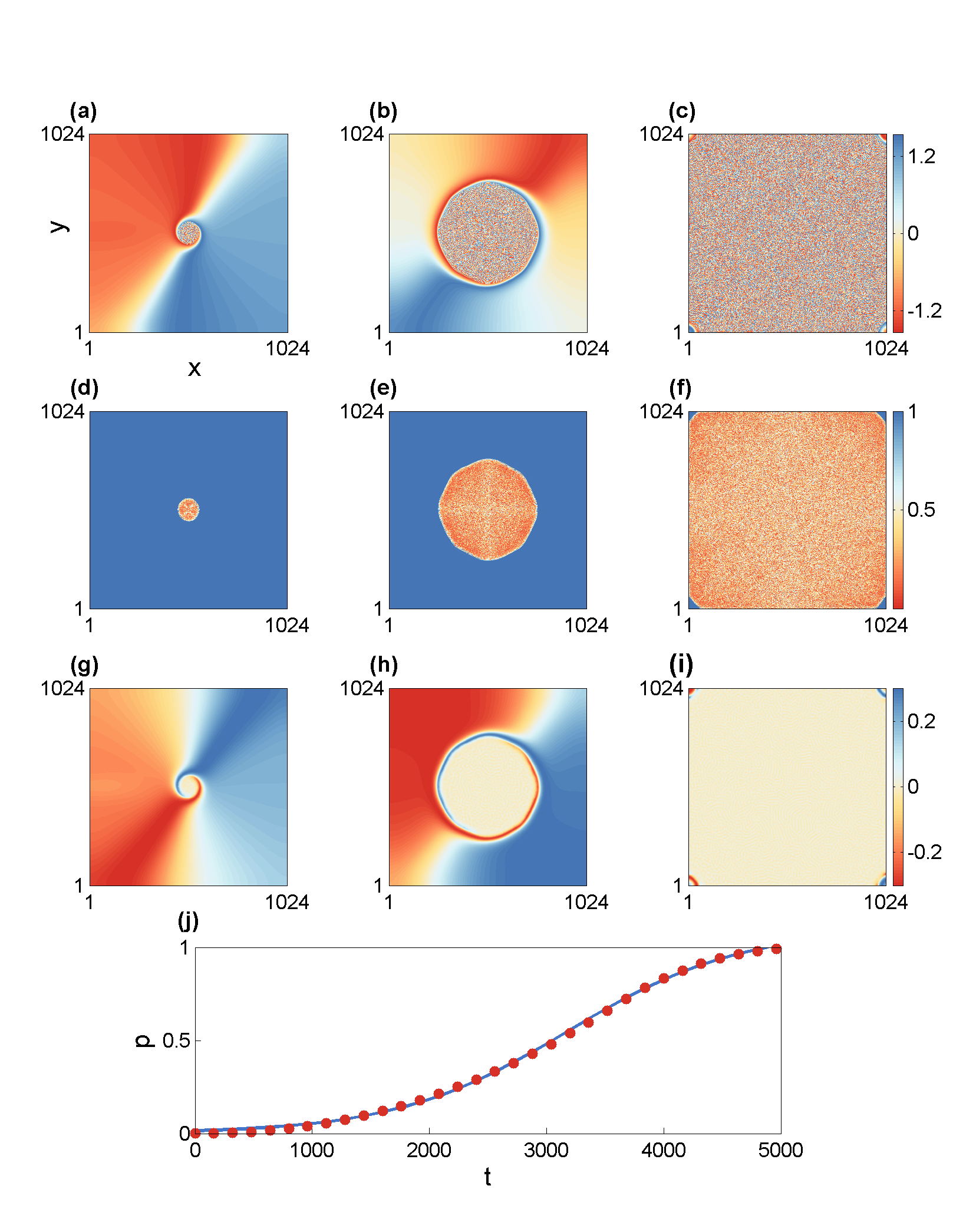}
\caption{Transition from SWC to completely incoherent state. The parameters are $a=3.0$ and $\epsilon_{2}=0.40$. (a-c) Typical states observed in the evolution of component $u$ at $t=500$, $2000$ and $5000$. (d-f) Typical states of the local order parameter $R$. (g-i) Typical states of the component $w$. (j) The time evolution of the fraction of asynchronous oscillators, $p$. Red dots: numerical results obtained by simulations. Blue curve: the fitted growth, which is identical to the one obtained for SWC in Fig. \ref{fig7}(j).}
\label{fig9}
\end{figure*}

To characterize the phenomenon of shadowed spirals further, we focus on the behaviors of the core area, and investigate the spatial distribution of the components $u$ and $w$ and the local order parameter $R$. The enlarged view of the core area for component $u$ is plotted in Fig. \ref{fig8}(a). Fixing $x=512$ in Fig. \ref{fig8}(a), we plot in Fig. \ref{fig8}(b) the distributions of $u$ (top) and its gradient $\Delta u$ (bottom), respectively. The top panel in Fig. \ref{fig8}(b) shows that, while $u$ is varying wildly along $y$, some regular patterns are found in the variation. To be more specific, the variation reaches its local maxima at a regular spatial distance $\lambda_{u}\approx 26$. However, when looking at the gradient $\Delta u$, the patterns found in $u$ are disappeared and the variation is completely random, as depicted in the bottom panel in Fig. \ref{fig8}(b). Figures \ref{fig8}(a) and (b) confirm the fact that spiral wave is only observable at large scale ($>\lambda_{u}$), while at small scales the system is completely disordered. The similar features are also observed for the local order parameter $R$, as depicted in Figs. \ref{fig8}(c) and (d). Comparing to the distribution of $u$, we see that in the top panel of Fig. \ref{fig8}(d) the distribution of $R$ is more irregular. The irregular variation of $R$ is understandable, as it represents the synchronization degree of five neighboring oscillators on the grid, which smears the patterns shown in the top panel of Fig. \ref{fig8}(b). The pattern of spiral waves is more prominent for the component $w$, as depicted in Figs. \ref{fig8}(e) and (f). The distribution of $w$ along the vertical central axis is plotted in the top panel of Fig. \ref{fig8}(f). We see that $w$ is varying in a regular manner along $y$, with the wave length being $\lambda_w\approx 26$ [identical to that of $u$ shown in the top panel of Fig. \ref{fig8}(b)]. The distribution of $\Delta w$ is plotted in the bottom panel of Fig. \ref{fig8}(f), which shows that $\Delta w\approx 0$ along the axis. We note that, while the similar phenomenon can be observed by introducing noise perturbations to the conventional spiral waves, the phenomenon of shadowed spirals reported here is emerged as a self-organization pattern of the locally coupled dynamical elements. That is, shadowed spirals is an inherent pattern of the system.

The expansion of the asynchronous core does not always lead to shadowed spirals. Varying the parameters $a$ and $\epsilon_2$, cases can be found in which the expansion of the asynchronous core leads to the development of completely incoherent state. (See Supplementary Materials for the movie.) An example of this is shown in Fig. \ref{fig9}, in which the parameters are $a=3.0$ and $\epsilon_{2}=0.4$. Typical states in the evolution of the component $u$ are shown in Figs. \ref{fig9}(a-c). We see that as time increases, the asynchronous core is expanding in size, and finally occupies the whole space. Yet, different from shadowed spirals, no pattern is observed in the asynchronous core in the transient states, neither in the finally state. That is, oscillators in the asynchronous core are completely desynchronized. The feature of desynchronized oscillators inside the core is shown more clearly in the evolution of the local parameter $R$, as depicted in Fig. \ref{fig9}(d-f). We see that $R<1$ inside the core while $R\approx 1$ outside. The evolution of the component $w$ is shown in Fig. \ref{fig9}(g-i). We see that, unlike the behaviors of $u$ and $R$, the distribution of $w$ is almost uniform. Comparing completely incoherent state with shadowed spirals, we see that the former might be regarded as the removal of the spirals from the latter. As a verification of this conjecture, we plot in Fig. \ref{fig9}(j) the variation of $p$, the fraction of asynchronous oscillators on the grid, with respect to time. The results are similar to that of shadowed spirals shown in Fig. \ref{fig7}(j).     

\begin{figure*}[tbp]
\center
\includegraphics[width=0.8\linewidth]{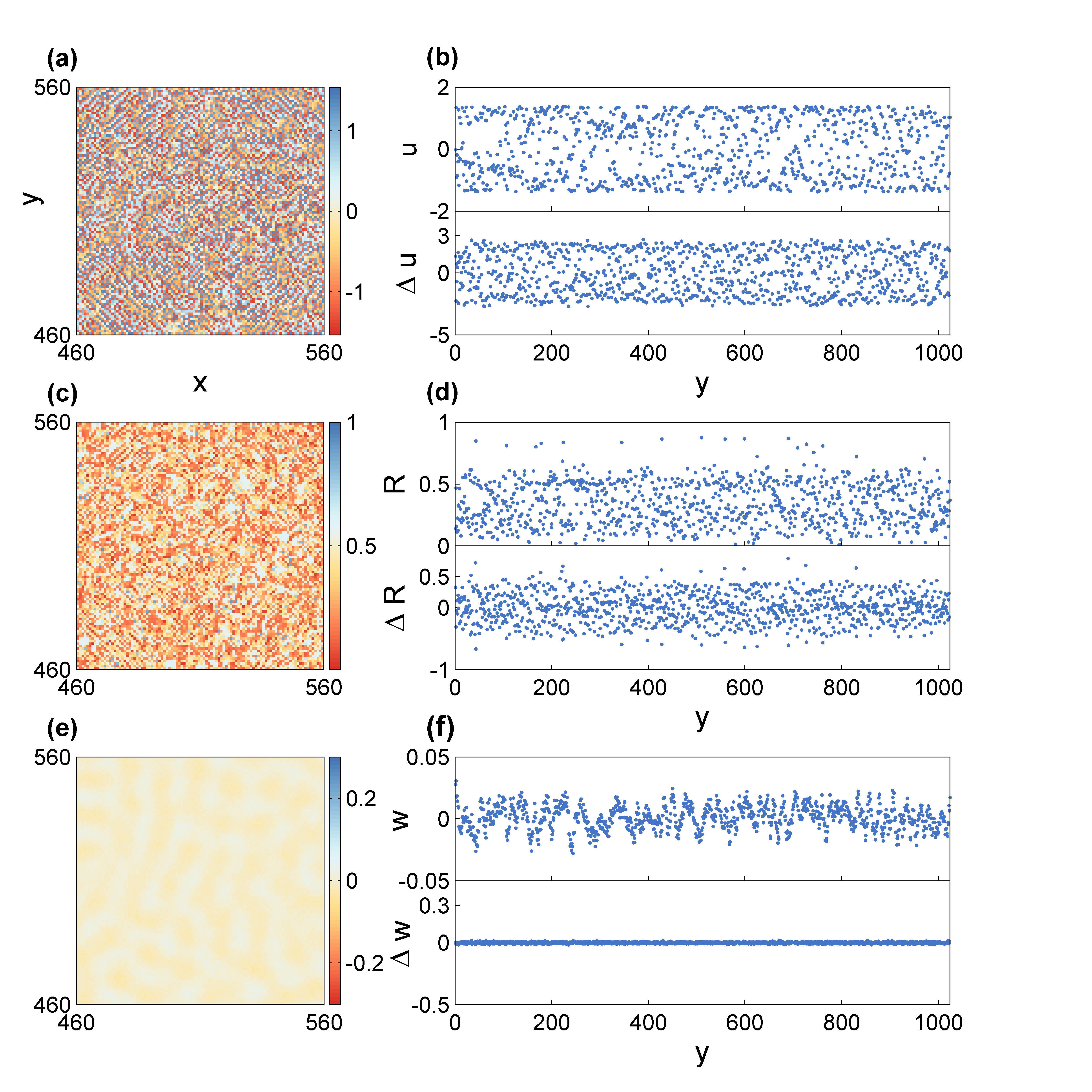}
\caption{Properties of completely incoherent state in the core area. (a) Enlarged view of component $u$. (b) Distributions of $u$ (top) and $\Delta u$ (bottom) along the central vertical axis ($x=512$) in (a). (c) Enlarged view of the local order parameter $R$. (d) Distributions of $R$ (top) and $\Delta R$ (bottom) along the central vertical axis. (e) Enlarged view of component $w$. (f) Distributions of $w$ (top) and $\Delta w$ (bottom) along the central vertical axis. Note that in the top panel the value of $w$ is fluctuating around $0$ with a very small amplitude ($\sim 5\times 10^{-2}$). The parameters are the same as in Fig. \ref{fig9}.}
\label{fig10}
\end{figure*}

To have more details on the properties the complete incoherent state, we plot in Fig. \ref{fig10} the distributions of $u$, $R$ and $w$ in the core region, together with their distributions along the vertical central line ($x=512$). Comparing the results with that of shadowed spirals [see Fig. \ref{fig8}], we see that in Fig. \ref{fig10} no regular pattern is observed in $u$, $R$ and $w$, neither in the variations of $\Delta u$, $\Delta R$ and $\Delta w$. In particular, the top panel of Fig. \ref{fig10}(f) shows that $w\approx 0$ for all sites, which is apparently different from the results of shadowed spirals [see the top panel in Fig. \ref{fig8}(f)]. The results in Fig. \ref{fig10} confirm the absence of any pattern in the core area, and imply that oscillators inside the core are completely desynchronized.

\subsection{The global picture}

\begin{figure}[tbp]
\center
\includegraphics[width=\linewidth]{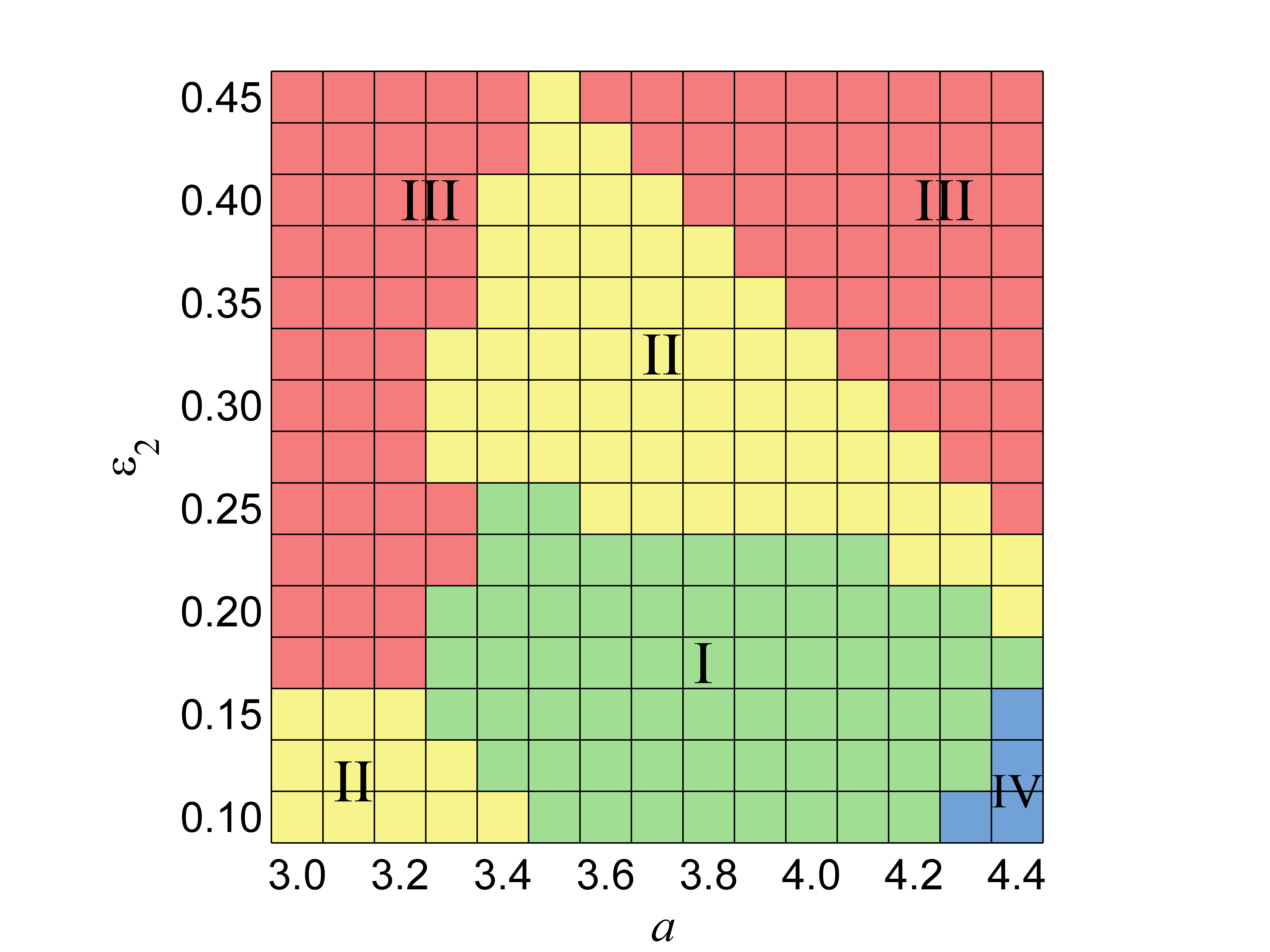}
\caption{Phase diagram in the parameter space of $(a,\epsilon_{2})$. Region I (green color): stable SWC. Region II (yellow color): SWC turbulence generated by core breakup. Region III (red color): shadowed spirals and completely incoherent states generated by core expansion. Region IV (blue color): other states observed in simulations.}
\label{fig11}
\end{figure}

To have a global picture on the bifurcation diagram, we check numerically the distribution of the final state in the parameter space of $(a,\epsilon_2)$. In simulations, the system is always initialized with the same conditions (as described in Sec. II), and the final state is taken at $t=2\times 10^5$ of the system evolution. The results are presented in Fig. \ref{fig11}. In Fig. \ref{fig11}, SWC is observed in region I (green color), SWC turbulence generated by core breakup is observed in region II (yellow color), shadowed spirals and completely incoherent states generated by core expansion are observed in region III (red color), and region IV (blue color) denotes the other states. Please note that region III is constituted by two disconnected small regions, one at the top left corner and the other one at the top right corner. Shadowed spirals and completely incoherent states are observed in both two small regions, and are entangled in the parameter space without a clear boundary. However, in terms of shadowed spirals, numerically we find that the wave length of the spirals generated in the top right region is much larger than that in the top left region. Figure \ref{fig11} reveals the rich dynamics inherent in the model of locally coupled RD system we have proposed, and provides a guideline for finding SWCs and shadowed spirals in simulations and experiments.

Besides the aforementioned states, other interesting states can be also emerged in the locally coupled RD system. These states are observed in region IV in Fig. \ref{fig11}. For instance, a new type of traveling wave containing an asynchronous strip is observed. As time increases, the plane waves are traveling inward, and are vanished after reaching the asynchronous strip. (See Supplementary Materials for the movie.) Different from the conventional traveling and target waves observed in RD systems, a fascinating feature of this state is that the source of the waves is composed of a group of desynchronized oscillators. As the oscillators in the strip are desynchronized, the new type of traveling wave thus is generated without the presence of periodic forcing or heterogeneity -- a necessary condition for generating traveling and target waves in conventional RD systems. A full exploration of the states in region IV is out the scope of the current study.

\section{Discussions and Conclusion}

Whereas chimera-like patterns have been reported in a variety of systems in literature, most of the studies rely on the adoption of nonlocal couplings. As nonlocal couplings are absent in typical RD systems in which elements are interacted through local diffusions, it is commonly believed that chimera-like pattens such as SWC can not be observed in typical RD systems. This believing is validated further by experiments in Ref. \cite{Totz_np18}. There, to generate SWC in chemical BZ oscillators, nonlocal couplings are established between remote oscillators through a computer interface. In contrast to the conventional wisdom, our present work shows that SWC can be generated in typical RD systems of local diffusion as well. Moreover, we demonstrate that by varying the system parameters, rich dynamics and bifurcation scenarios can be presented in the proposed system. In particular, a new phenomenon named shadowed spirals is observed, in which regular spirals are embedded in the background of globally desynchronized oscillators. This new phenomenon, which to the best of our knowledge has not been reported in literature, manifests from a new viewpoint the coexistence of incoherent and coherent states in spatially extended systems, generalizing thus the traditional concept of chimera states. Besides offering rich dynamics, the system presents also clear scenarios on the transitions of the system dynamics. Specifically, two different scenarios have been revealed in the destabilization of SWCs: core breakup and core expansion. While similar scenarios have been reported in Ref. \cite{Totz_np18}, the results in Ref. \cite{Totz_np18} are obtained in nonlocally coupled oscillators. Moreover, we conduct in the present work a detailed analysis on scenarios of SWC transitions and, more importantly, explore the bifurcation diagram of SWC in the two-dimensional parameter space, which deepens our understandings on the nature of SWCs. As nonlocal couplings are employed in Ref. \cite{Totz_np18} and the present work utilizes local couplings, our study thus suggests that the above scenarios might be universal for the transitions of SWC in RD systems. 

The theoretical model we have proposed could be realized in experiments. As analyzed in Sec. \ref{mechanism}, the key to generating SWC lies in the single-component diffusion of the chemical reactants, i.e., diffusion exists only for the component $w$, while is absent or very weak for components $u$ and $v$. The essence of such a setting is to separate the diffusive component from the non-diffusive ones, protecting thus the asynchronous cores of the latter from diffusion-induced destructions. A candidate system approximately satisfying the requirement of single-component diffusion is the BZ-AOT systems \cite{Cherkashin_jcp08}, in which the diffusion coefficients of the activator species ${\rm HBrO_{2}}$ (water soluble) and the inhibitors ${\rm Br^{-}}$ (water soluble) are much smaller to that of the inhibitors ${\rm Br_{2}}$ (oil soluble). Single-component diffusion can be also realized in synthetic biological systems, saying, for instance, the genetic regulation network of {\it Escherichia coli} cells \cite{Danino_nat10}, the yeast cell layers \cite{Shutz_BJ11}, and the {\it Dictyostelium} cells \cite{Noorbakhsh_pre15}. In these systems, the dynamical elements are not interacting directly, but through a common diffusive environment. From the point of view of indirectly coupled oscillators, these systems share the same nature of the model we have proposed, and therefore are also suitable candidates for generating SWCs. 

In summary, we have proposed an experimentally feasible model of locally coupled RD system, and investigated the formation of SWC and the transitions from SWC to other states. The conditions for generating SWC have been given, and the underlying mechanism has been analyzed by a phenomenological theory. The transitions from SWC to other states in the two-dimensional parameter space of experimental interest have been studied, which shows that SWC is typically destabilized through two scenarios, core breakup and core expansion. Details of the transition processes have been explored, and new patterns have been observed. In particular, a new type of chimera, namely shadowed spirals, has been uncovered. A global picture of the bifurcation diagram has been given, which shows that, in contrast to the conventional wisdom, SWC can be generated in a wide region in the parameter space of the locally coupled RD system. Our studies shed new lights on the formation of chimera states in spatiotemporal systems, and pay the way for generating SWC in typical RD systems in experiments.

\section*{Acknowledgments}

 This work was supported by Natural Science Foundation of Zhejiang Province under Grant No. LY16A050003. XGW was supported by the National Natural Science Foundation of China under Grant No. 11875182.

\end{document}